\documentclass[pre]{revtex4}
\usepackage{setspace}
\usepackage{graphicx}
\usepackage{dcolumn}
\usepackage{amsmath}
\usepackage{amsfonts}
\usepackage{amssymb}
\setcitestyle{super}

\begin{document}

\title{Solvent-quality dependent contact formation dynamics in proteins}
\author{Prasanta Kundu and Arti Dua}
\affiliation{Department of Chemistry, Indian Institute of Technology, Madras, Chennai-600036, India}

\date{\today}

\begin{abstract}
\noindent
The mean time of contact formation between two ends of a protein chain shows power law dependence with respect to the number of residues, $\tau_{CF} \sim N^{\alpha}$. Fluorescence quenching measurements based on triplet-triplet energy transfer show variation in the value of scaling exponent $\alpha$ for different protein-solvent systems. This points to the relevance of the protein-solvent interactions (solvent quality) and hydrodynamic interactions in determining the time scale of contact formation. Here, starting from a non-Markovian diffusion equation supplemented with an exponential sink term that accounts for the energy transfer reaction between donor and acceptor groups, we calculate the mean time of contact formation using the Wilemski-Fixman closure approximation. The non-Markovian diffusion-reaction equation includes the effects of solvent quality and hydrodynamic interaction in a mean-field fashion. It shows that the contact formation dynamics is mainly governed by two time scales, the reciprocal of the intrinsic rate of quenching $(k_0^{ET})^{-1}$, and the relaxation time $\tau_0 = \eta b^3/k_B T$ of the coarse-grained residue of an effective size $b$ with solvent viscosity $\eta$. In the limit of $k_0^{ET} \tau_0 \ll 1$, the dominating effect of the reaction-controlled kinetics yields the scaling exponents as $0.89$, $1.47$ and $1.79$ in poor, theta and good solvents respectively. In the opposite limit $k_0^{ET} \tau_0 \gg 1$, the dominating influence of the diffusion-controlled kinetics results in $\alpha$ as $1.90$, $2.17$, $2.36$ for a freely-draining and $1.31$, $1.77$, $2.06$ for a non-freely-draining chain in poor, theta and good solvents respectively. In the intermediate limit,  $k_0^{ET} \tau_0 \approx 1$, the increase in the number of residues switches the kinetics from reaction-controlled at low $N$ to diffusion-controlled at large $N$. These general results suggest that experimental estimates of the scaling exponents reflect solvent-quality dependence of the mean contact formation time in the reaction-controlled limit.
\end{abstract}
\maketitle

\pagebreak

\section{Introduction}
\noindent

The rate at which protein conformational space can be explored to form intramolecular contact between two residues in a polypeptide chain  is an important  elementary process in protein folding.$\cite{Eaton2000}$ In the past few years, triplet-triplet energy transfer and photoinduced electron transfer between donor and acceptor groups located at two ends of a polypeptide chain have offered novel means to probe the role of intramolecular chain diffusion in  determining the rate of contact formation and the time scale of protein conformational fluctuations.\cite{Lapidus2000, Lapidus2002, Bieri1999, Fierz2007, Krieger2003, Chang2003, Hudgins2002, Lee2007, Ford2013, Neuweiler2003} By monitoring the triplet-triplet absorption, these experiments have obtained the effective rate constant $k_{CF}$ for contact formation between two ends of a protein chain, the reciprocal of which is equal to the mean time of contact formation, $\tau_{CF} = k_{CF}^{-1}$. For $N \simeq 10-20$, the latter follows a power law dependence with respect to the number of residues, $\tau_{CF} \propto N^{\alpha}$, where $\alpha$ is the scaling exponent. The typical values of  $\alpha$  for different protein-solvent systems are  $1.05\pm0.06$,\cite{Hudgins2002} $1.5$\cite{Lapidus2002,Buscaglia2003} and $1.7\pm0.1$\cite{Krieger2003, Fierz2005}  for $N \simeq 10-20$. For relatively shorter polypeptides, where effects of stiffness become important, the dependence of $\tau_{CF}$ on $N$ is much weaker.\cite{Lapidus2000, Santo2009}

In the asymptotic limit of large $N$, when the reaction between two end residues of a freely-draining chain in a theta solvent is considered instantaneous, the theoretical estimates of diffusion-controlled mean time of contact formation based on the Wilemski-Fixman (WF) and Szabo-Schulten-Schulten (SSS) formalisms yield  $\tau_{wf}^{d} \propto N^2$ and $\tau_{sss}^{d} \propto N^{3/2}$ respectively.\cite{Wilemski1974, Szabo1980} The simplicity of the SSS formalism and  its close agreement with the experimental scaling exponent of $1.5$ makes it a widely used theory to rationalize experimental data on end-to-end contact formation in polypeptides. The SSS prediction, however, deviates both from the WF theory\cite{Wilemski1974} and simulation results.\cite{Pastor1996, Sakata1976, Podtelezhnikov1997, Chen2005} In a recent work on contact formation kinetics\cite{Toan2008}, it is shown that the SSS formalism also yields $\tau_{esss}^d \propto N^2$ once the monomer diffusion coefficient is replaced with an effective diffusion coefficient that includes the relaxation dynamics of the chain ends. The extended-SSS theory\cite{Toan2008} yields $\alpha \sim 2$ in agreement with the WF formalism, but can not rationalize the weaker dependence ($\alpha < 2$) of $\tau_{CF}$ on $N$ observed experimentally. 


The WF formalism determines  the mean time of contact formation by solving the reaction-diffusion equation in the presence of a sink term. The sink term accounts for the probability of end-to-end  contact whenever the ends are within a contact distance $a$. For an idealized sink given by $k_I(R) = k_I^0 \delta(R-a) $, the reaction between two ends is instantaneous as soon as  $R = a$. The diffusion-controlled (DC) limit of $k_I^0 \rightarrow \infty$, thus, yields $\tau_{wf}^d \sim \tau_0 N^{2}$ as the mean time of contact formation between two ends of a freely-draining ideal polymer.\cite{Chakrabarti2012}  For measurements based on triplet-triplet energy transfer, however, the quenching  rate depends exponentially on the distance between the donor and acceptor.  The latter is given by  $k_{ET} (R) = k_0^{ET} \exp(-2 R/a)$, where $k_0^{ET}$ is the intrinsic quenching rate independent of the distance between donor and acceptor groups $R$ and $a$ is the contact distance for quenching. In the presence of a more realistic energy transfer sink, therefore, the reaction is not instantaneous but occurs at a rate that decays exponentially with $R$. Typical values of $k_0^{ET}$, range from $10^6-10^9 s^{-1}$.\cite{Lapidus2000} For large but finite $k_0^{ET}$, it is conceivable that the weaker dependence of $\tau_{CF}$ on $N$  can be rationalized using the reaction-controlled kinetics. Additionally, the presence of protein-solvent interactions which effectively amounts to change in the solvent quality  from theta to good or poor solvent conditions, and  hydrodynamic interaction which couples the dynamics of various residues in the chain can result in weaker dependence of $\tau_{CF}$ on  $N$.\cite{Lapidus2002}

In this work, starting from a non-Markovian diffusion equation supplemented with an exponential sink term that accounts for the energy transfer reaction  between two residues located at the exterior of a protein chain, we calculate the mean time of contact formation using the WF closure approximation.\cite{Wilemski1974, Dua2002} The non-Markovian diffusion equation describes the time evolution of the probability distribution of the distance between two residues on a protein chain and includes the effects of solvent quality and hydrodynamic interaction in a mean-field fashion. Our key result is that for  triplet-triplet energy transfer, where the quenching rate depends exponentially on the distance between donor and acceptor groups, the kinetics of contact formation is reaction-controlled (RC) in the limit of $k_0^{ET} \tau_0 \ll 1$ and diffusion-controlled (DC) in the opposite limit, $k_0^{ET} \tau_0 \gg 1$. Here, $\tau_0 = \eta b^3/k_B T$ is the relaxation time of the coarse-grained residue of an effective size $b$ (molecular relaxation time).  In the intermediate limit,  $k_0^{ET} \tau_0 \approx 1$, the increase in the number of residues switches the kinetics from reaction-controlled at small $N$ to diffusion-controlled at large $N$.  Our analysis shows that even for large values of $k_0^{ET}$, the weaker dependence of $\tau_{CF}$ on $N$, can be rationalized using reaction-controlled kinetics in poor, theta and good solvent conditions. In the presence of the heaviside sink, the diffusion-controlled mean time of contact formation obtained from the present formalism is in agreement with the previous work based on generalized random walk description that accounts for non-local interactions approximately.\cite{Debnath2004}

This paper is organized as follows. Section II recapitulates the general features of a non-Markovian generalized Langevin equation (GLE), which is modified to include the effects of solvent quality and hydrodynamic interaction in a mean-field fashion. The GLE, when transformed into a diffusion equation and supplemented with an exponential sink term, results in a non-Markovian diffusion-reaction equation. In Section III, the latter is used to determine the mean time of contact formation using the WF closure approximation. Section IV presents the main results of this calculation along with a brief discussion. Conclusions are presented in Section V.

\section{Theoretical model}
The theoretical model presented  below is based on a recent work where the time evolution of distance between donor and acceptor groups on a protein chain was described using an overdamped non-Markovian generalized Langevin equation (GLE) approach.\cite{Dua2011} This model was used to rationalize the results of a recent experiment where fluorescence quenching of photoinduced electron transfer between a pair of donor and acceptor groups on a protein chain was used to probe several universal aspects of protein conformational fluctuations.\cite{ Kou2004} In this experiment, distance fluctuations were shown to follow the Gaussian statistics with non-exponential decay, revealing non-Markovian nature of these fluctuations. The GLE  approach\cite{Dua2011} captured several universal aspects of the photoinduced electron transfer experiment, including correct prediction of the power law for the memory kernel\cite{Min2005} and excellent agreement with two-point and four-point fluorescence correlation lifetimes.\cite{ Kou2004}  Here, we modify the GLE to include the effects of solvent quality and hydrodynamic interactions in a mean-field fashion:
\begin{equation}\label{gle}
\int_0^t ~dt^{\prime} ~K_{mn}(t-t^{\prime}) \frac{d}{dt^{\prime}} {\bf R}_{mn}(t^{\prime}) = - \kappa {\bf R}_{mn}(t) + {\bf f}_{mn}(t).
\end{equation}
In the above equation ${\bf R}_{mn} = {\bf r}_n - {\bf r}_m$ is the distance between two residues labelled as $m$ and $n$ and  located at positions ${\bf r}_m$ and ${\bf r}_n$ respectively on the protein chain. The first term on the right hand side is the effective elastic force due to chain connectivity with free energy $F({\bf R}_{mn}) = \frac{1}{2}\kappa {\bf R}_{mn}^2$, where $\kappa = 3k_B T/\left<{\bf R}_{mn}^2\right> $ is the force constant.\cite{Doi1986} The variation in the solvent quality from theta to good and poor solvent conditions results in effective repulsive and attractive interactions between chain residues respectively. Within a mean-field description, the latter can be included by considering $ \left<{\bf R}_{mn}^2\right> = |n-m|^{2\nu} b^2$, where $\nu$ is the Flory exponent with $\nu = 1/2$, $3/5$ and $1/3$ for theta, good and poor solvent conditions respectively.\cite{Doi1986, Rubinstein2003} Thus the above GLE, while retaining the Gaussian and non-Markovian nature of distance fluctuations between two residues, accounts for the effective non-local interactions between chain residues in an approximate manner.

In Eq. (\ref{gle}), the mean and the variance of the Gaussian coloured noise ${\bf f}_{mn}$ are given by $\langle {\bf f}_{mn}(t)\rangle = 0$ and $\langle f_{mn}(t)f_{mn}(0)\rangle = k_BT K_{mn}(t)$ respectively, where $K_{mn}(t)$ is the friction kernel. The Laplace transform of $K_{mn}(s) = \int_0^{\infty}dt \exp(-st) K_{mn}(t)$ is given by
\begin{equation}\label{kernel}
K_{mn}(s) =  \frac{ \kappa \phi_{mn}(s)}{ 1 -  s \phi_{mn}(s)}, 
\end{equation}
where $\phi_{mn}(s) =\int_0^{\infty}dt \exp(-st) \phi_{mn}(t)$ is the dimensionless time correlation of distance fluctuations. The latter can be obtained from the Rouse and Zimm dynamics in theta, good and poor solvents. An outline of the derivation is presented in Appendix A. The final expression is
\begin{equation}\label{corr1}
\phi_{mn}(t) = \frac{\left< {\bf R}_{mn}(t)\cdot {\bf R}_{mn}(0) \right>}{\left< {\bf R}_{mn}(0) \cdot {\bf R}_{mn}(0) \right>} =  \frac{\sum_{p=1}^{\infty}  [\cos(p\pi n/N) - \cos(p\pi m/N)]^2 e^{- t/\tau_{p}}/p^{2\nu+1}}{\sum_{p=1}^{\infty}  [\cos(p\pi n/N) - \cos(p\pi m/N)]^2/p^{2\nu+1}},
\end{equation}
where $N$ is the total number of residues in the protein chain and $\tau_p$ is the relaxation time of the pth mode.. The above equation accounts for the solvent quality and hydrodynamic interaction in a mean-field fashion. For freely-draining and non-freely-draining chain corresponding to the absence and presence of hydrodynamic interaction, the relaxation time is specified using $\tau_p^R$ and $\tau_p^Z$, where superscripts $R$ and $Z$ refer to the Rouse and Zimm dynamics respectively. In terms of the longest relaxation time corresponding to the $p=1$ mode, $\tau_p^R = \tau_1^R/p^{2\nu + 1}$ and $\tau_p^Z =\tau_1^Z/p^{3\nu}$, where $\tau_1^R \sim  \tau_0 N^{2\nu+1}$ and $\tau_1^Z \sim \tau_0 N^{3\nu}$ respectively. Here,  $\tau_0 \sim \eta b^3/ k_B T$ is the relaxation time of the coarse-grained residue of size $b$ (Kuhn length) with solvent viscosity $\eta$.\cite{Doi1986, Rubinstein2003}

Eq. (\ref{gle}) when transformed into a Smoluchowski equation is given by
\begin{equation}\label{smolu}
\frac{\partial P(\textbf{R}_{mn}, t)}{\partial t} =   
D_{mn}(t) \left[\frac{\partial}{\partial \textbf{R}_{mn}} \cdot \,{\textbf{R}_{mn}~P(\textbf{R}_{mn},t)} + \frac{k_B T}{\kappa}  \frac{\partial^2}{\partial \textbf{R}_{mn}^2}   P(\textbf{R}_{mn},t)\right].
\end{equation}
In an earlier work\cite{Kundu2013} the survival probability of the unreacted donor state calculated from Eq. (\ref{smolu}) yielded excellent agreement with another recent experiment\cite{Wang2007} measuring temporal decay of the transient absorption signals for fourteen mutants and wild type reaction center of protein dynamics modulated electron transfer reaction in early stage of photosynthesis. Starting from the GLE [Eq. (\ref{gle})], a brief outline of the steps involved in deriving Eq. (\ref{smolu}) are presented in Ref. (30). 

Eq. (\ref{smolu}) is a non-Markovian diffusion equation, which describes the time evolution of the probability distribution of the distance between two residues on  a protein chain with time dependent diffusion coefficient, $D_{mn}(t)=-\frac{{\dot{\phi}}_{mn}(t)}{\phi_{mn}(t)}$. In the presence of an energy transfer reaction between donor and acceptor groups at the exterior (or interior) of the protein chain, Eq. (\ref{smolu}), is supplemented with an energy transfer sink term, resulting in the following diffusion-reaction equation:
\begin{eqnarray}\label{dre}
\frac{\partial P(\textbf{R}_{mn}, t)}{\partial t} &=&   
D_{mn}(t) \left[\frac{\partial}{\partial \textbf{R}_{mn}} \cdot \,{\textbf{R}_{mn}~P(\textbf{R}_{mn},t)} + \frac{k_B T}{\kappa}  \frac{\partial^2}{\partial \textbf{R}_{mn}^2}   P(\textbf{R}_{mn},t)\right] - k_{ET}({R}_{mn}) P(\textbf{R}_{mn},t) 
\end{eqnarray}
Triplet-triplet energy transfer follows Dexter electron exchange as a mechanism for fluorescence quenching.\cite{Valeur2001} Thus, the distance dependent energy transfer rate expression is given by
\begin{equation}\label{etr}
k_{ET}(R_{mn})  = k_0^{ET} \exp\left(- 2 R_{mn}/a\right),
\end{equation}
where $k_0^{ET}$ is the intrinsic rate constant which depends on the spectral overlap integral and $a$ is the contact distance for quenching. In what follows, we use Eq. (\ref{dre}) to calculate the mean time of contact formation in the presence of the energy transfer reaction sink [Eq. (\ref{etr})].

\section{The mean time of contact formation}

The mean time of contact formation can be obtained from the survival probability $S(t) = \int d{\bf R}_{mn} P({\bf R}_{mn},t)$ that the donor and acceptor groups on the chain have not reacted at time $t$. From Eq. (\ref{dre}), it follows that $\frac{d S(t)}{dt} = - \left< k(t) \right>$, where $\left< k(t) \right> = \int_{-\infty}^{\infty}  ~d{\bf R}_{mn} ~ k_{ET}(R_{mn}) ~ P({\bf R}_{mn}, t)$.  If $S(t)$ is assumed to decay as a single exponential, $S(t) \approx \exp(- t/\tau_{CF})  =\exp(-k_{CF} t)  $, then the mean time of contact formation between two residues, which is the reciprocal of the effective rate constant for contact formation, is given by 
\begin{equation}
\tau_{CF} = k_{CF}^{-1}  = \int_0^{\infty} dt S(t)
\end{equation}

The solution of the diffusion-reaction equation, Eq. (\ref{dre}) can be written as\cite{Wilemski1974, Dua2002} 
\begin{equation}\label{dre-sol}
P({\bf R}_{mn}, t) = P_{eq}({\bf R}_{mn}) - \int_0^{t} dt' \int_{-\infty}^{\infty}  d{\bf R}_{mn}' G({\bf R}_{mn}, t | {\bf R}_{mn}', t') k_{ET}({R}_{mn}') P({\bf R}_{mn}', t')
\end{equation} 
where $G({\bf R}_{mn}, t | {\bf R}_{mn}', t')$ is the conditional probability that the distance between two residues on a protein chain which was ${\bf R}_{mn}'$ at  some initial time $t'$ is ${\bf R}_{mn}$ at time $t$. Starting from Eq. (\ref{smolu}), a closed form expression for $G({\bf R}_{mn}, t | {\bf R}_{mn}', t')$ can be derived, the details of which are given in Ref. (30). Here, we simply state the final result: 
\begin{eqnarray}\label{gaussian}
G\left({\bf R}_{mn}, t | {\bf R}_{mn}', 0 \right) =\left( \frac{3}{2 \pi |n-m|^{2\nu} b^{2} \left[1-\phi_{mn}^{2}(t)\right]} \right)^{3/2} 
\exp \left[- \frac{3 \left({\bf R}_{mn}- {\bf R}_{mn}' \phi_{mn}(t)\right)^{2}}{2 |n-m|^{2\nu} b^{2} \left[1-\phi_{mn}^{2}(t)\right]} \right].
\end{eqnarray}
From the above equation, the following equilibrium distribution can be obtained in the long time limit, $t \rightarrow \infty$:
\begin{equation}\label{ssgaussian}
P_{eq}({\bf R}_{mn}) =\left( \frac{3}{2 \pi |n-m|^{2\nu} b^{2} }\right)^{3/2} 
\exp \left[- \frac{3 {\bf R}_{mn}^2}{2 |n-m|^{2\nu} b^{2}} \right].
\end{equation}
which correctly yields $\left<{\bf R}_{mn}^2\right>_{eq} = |n-m|^{2\nu} b^{2}$ for the distance between two residues at the interior of the chain and $\left<{\bf R}^2\right>_{eq} = N^{2\nu} b^{2}$ for the end-to-end distance with $n=N$ and $m=0$. 

Eq. (\ref{dre-sol}), when multiplied with $k_{ET}(R_{mn})$ and integrated over $d{\bf R}_{mn}$ yields
\begin{equation}\label{dre-sol1}
\left< k(t) \right> = \left< k \right>_{eq} - \int_0^{t} dt' \int_{-\infty}^{\infty}  d{\bf R}_{mn} \int_{-\infty}^{\infty}  d{\bf R}_{mn}' ~ k_{ET}(R_{mn}) ~G({\bf R}_{mn}, t | {\bf R}_{mn}', t')~ k_{ET}({R}_{mn}') ~P({\bf R}_{mn}', t')
\end{equation}
where $\left< k \right>_{eq} =  \int_{-\infty}^{\infty}  ~ d{\bf R}_{mn} ~ k_{ET}(R_{mn}) ~ P_{eq}({\bf R}_{mn})$. Eq. (\ref{dre-sol1}) is a nonlinear integral equation which can not be solved in a closed form. Using Wilemski-Fixman closure approximation we determine the mean time of contact formation, the derivation of which is discussed at length in Ref. (23). The final result is 
\begin{eqnarray}\label{skin}
\tau_{CF} = k_{CF}^{-1} &=& \left< k \right>_{eq}^{-1} + \int_0^{\infty}dt~ \frac{(\left< k(t) k(0)\right>_{eq} - \left< k \right>_{eq}^2)}{\left< k \right>_{eq}^2}\nonumber\\
&= & \tau_R + \tau_D,
\end{eqnarray}
where 
\begin{equation} 
\left< k(t) k(0)\right>_{eq} =  \int_{-\infty}^{\infty} d{\bf R}_{mn}  \int_{-\infty}^{\infty} d{\bf R}_{mn}'  ~ k_{ET}(R_{mn}) ~G({\bf R}_{mn}, t | {\bf R}_{mn}', 0)~ k_{ET}({R}_{mn}') P_{eq}({\bf R}_{mn}'),
\end{equation}
is the time-correlation of the energy-transfer rates. In Eq. (\ref{skin}), $\tau_R$ and $\tau_D$ are the reaction and diffusion-controlled contributions to the mean time of contact formation respectively. For end-to-end contact formation, with $n=N$ and $m=0$,  the $\tau_R$ and $\tau_D$ are given by
\begin{eqnarray}\label{keq1}
\tau_R = \left< k \right>_{eq}^{-1}  &=& \left( 4 \pi k_0^{ET} \left( \frac{3}{2 \pi N^{2\nu} b^{2} }\right)^{3/2}  \int_a^{\infty} ~ d R ~R^2 \exp(- 2 R/a )~g(R) \right)^{-1}.
\end{eqnarray}
and
\begin{eqnarray}\label{kcorr1}
\tau_D &=&  \int_0^{\infty} dt \left[\frac{\left< k(t) k(0)\right>_{eq}}{\left< k \right>_{eq}^2} - 1 \right] \nonumber\\
&=& \int_0^{\infty} dt \left[ \frac{N^{2\nu} b^2}{3 \phi(t) \left[1-\phi^{2}(t)\right]^{1/2} } \frac{ \int_a^{\infty} ~ d R \int_a^{\infty} ~ d R' ~R ~R'  ~ e^{- 2 (R+R')/a} ~ f(R, t) }{\left(\int_a^{\infty} ~ d R ~R^2~ e^{- 2 R/a}~g(R)\right)^2} - 1 \right]\nonumber\\
\end{eqnarray}

\begin{figure}[t]
\centering
\includegraphics[trim=0.0cm 1.5cm 0cm 1cm,clip=true,scale=0.7]{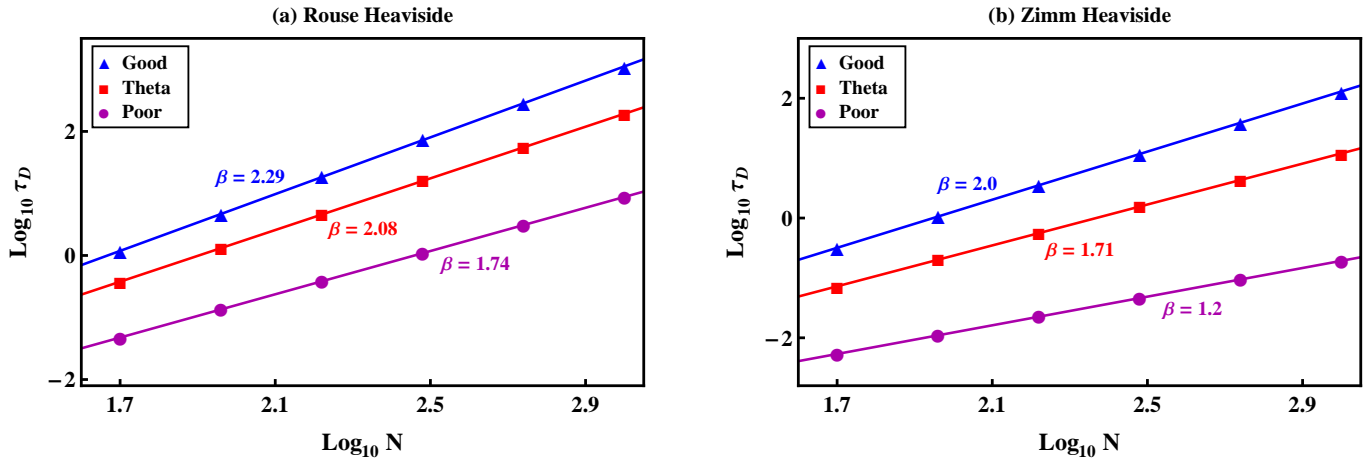}
\caption{Log-log plot for the variation of the diffusion-controlled mean time of contact formation $\tau_D$ with $N$ for a chain in good, theta  and poor solvents. The results are obtained in the presence of the heaviside sink in the (a) absence (Rouse) and (b) presence (Zimm) of hydrodynamic interaction. Symbols are theoretical results and the solid lines are the corresponding linear fits showing the power law dependence of diffusion controlled mean time of contact formation with respect to the number of residues, $\tau_D \sim N^{\beta}$.}
\end{figure} 

where $f(R, t) =  \exp{\left[-\frac{3( {R}^2 + R'^2)}{(2 N^{2\nu} b^{2} (1-\phi^{2}(t)))}\right]} \sinh\left[\frac{3 R R' \phi(t)}{ N^{2\nu} b^{2} (1-\phi^{2}(t))}\right] $, $g(R) =  \exp{ \left[-\frac{3 {R}^2}{2 N^{2\nu} b^{2}}\right]}$. In the above equation, $\phi(t)$ is given by
\begin{equation}
\phi(t) = \frac{\sum_{p, odd} e^{-t/\tau_p}/p^{2\nu + 1}}{\sum_{p, odd} 1/p^{2\nu + 1}},
\end{equation}
where $\tau_p^R = \tau_1^R/p^{2\nu+1}$ and $\tau_p^Z = \tau_1^Z/p^{3\nu}$ for the freely-draining and non-freely-draining chain respectively. In the next section, we numerically estimate $\tau_R$ and $\tau_D$ to obtain the mean time of contact formation using Eq. (\ref{skin}).

\section{Results and Discussion}

\subsection{Diffusion-controlled mean time of contact formation with the heaviside sink}

In a previous theoretical study, the solvent quality dependence of the mean contact formation time has been obtained using a generalized random walk description which accounts for the non-local interactions in a freely-draining chain approximately.\cite{Debnath2004} The nonlocal interactions have been included by modifying the connectivity term in the Edwards continuum representation of the polymer. This involves introducing a parameter $h$, with values 1/3, 1/2 and 3/5, which correspond to the average size of the chain in poor, theta and good solvents respectively. By solving the reaction-diffusion equation with the heaviside sink, the diffusion-controlled mean time of contact formation was shown to follow a power law scaling $\tau_D \propto N^{\beta}$.  The third and fourth columns of Table 1 present the values of the scaling exponent $\beta$ obtained in this previous study\cite{Debnath2004} and in a recent Brownian dynamics simulations\cite{Toan2008} respectively. These values have been obtained in good, theta and poor solvents in the absence of the hydrodynamic interactions (freely-draining chain).

\begin{table}[t]
\begin{center}
\centering
\begin{tabular}{c c c c c}

\hline \hline \\

           
  Solvent quality~ \hspace{0.5cm} & Present Work$^{a}$~~  \hspace{0.5cm}  &  Previous Work$^{1,a}$~~  \hspace{0.5cm}  & Previous Work$^{2,a}$~~   \hspace{0.5cm}  & Present Work$^{b}$ \hspace{0.5cm}  \\
   
                       &                        &                       &            &          \\  \hline
      
   \vspace{-.1cm} \\ [1ex]
   
Good \hspace{0.5cm} &  2.29  \hspace{0.5cm} & 2.28  \hspace{0.5cm}   & 2.4  \hspace{0.5cm}   & 2.0  \hspace{0.5cm}    \\ [1ex]

Theta \hspace{0.5cm} &  2.08  \hspace{0.5cm} & 2.09   \hspace{0.5cm}  & 2.0 \hspace{0.5cm}  & 1.71  \hspace{0.5cm}   \\ [1ex]
                                                                                     
Poor \hspace{0.5cm} & 1.74  \hspace{0.5cm}  & 1.76   \hspace{0.5cm}  & 1.0  \hspace{0.5cm}  & 1.2  \hspace{0.5cm} \\ [1ex]

%
%
      
\hline \hline
\end{tabular}

\end{center}
\caption{Comparison of solvent-quality dependence of scaling exponents $\beta$ for diffusion controlled mean time of contact formation $\tau_D \sim N^{\beta}$ in the presence of the Heaviside sink between present and previous$^{1}$ work. Superscripts $a$ and $b$ refer to the solvent quality dependence of the freely-draining and non-freely-draining chain corresponding to the absence (Rouse) and  presence (Zimm) of hydrodynamic interaction respectively. Superscripts $1$ and $2$ refer to previous estimates of the scaling exponents based on analytical theory and simulations for the freely-draining chain given in References [24] and [21] respectively.}
\end{table}

To compare the scaling exponents obtained in the present work with the earlier work, we first obtain the diffusion-controlled mean time of contact formation in the presence of the heaviside sink in good, theta and poor solvents. This is done by replacing $k_{ET}(R)$ in Eq. (\ref{dre}) with the heaviside sink given by  $k_{HS}(R) = k_{HS}$ for $R \leq a$ and $0$ otherwise. The diffusion-controlled mean time of contact formation is calculated in the limit $k_{HS} \rightarrow \infty$. In the absence and presence of hydrodynamic interaction corresponding to the  freely-draining and non freely-draining cases respectively it shows a power law scaling $\tau_D \propto N^{\beta}$, which is depicted in Fig. (1).  The values of the scaling exponents in good, theta and poor solvents in the absence of hydrodynamic interactions, obtained using this approach, are  tabulated in the second column of Table 1. These values show excellent agreement with the previous analytical results [column 3]. The values of  the scaling exponents obtained from simulations in good and poor solvents [column 4] are comparatively higher and lower than the previous and present analytical estimates. In Ref. (21), this has been attributed to finite size effects.

The comparison of the diffusion-controlled mean time of contact formation obtained from the present and previous\cite{Debnath2004}  theoretical approach suggests that  the non-Markovian diffusion equation approach, where the solvent quality is accounted for in a mean-field fashion, yields the same result as the generalized random walk description used earlier. However, the advantage of the present formalism is that it can easily be extended to account for the hydrodynamic interactions between chain residues. The scaling exponents for the diffusion-controlled mean time of contact formation in the presence of the hydrodynamic interactions are presented in column 5 of Table 1. In the presence of the hydrodynamic interaction, the scaling exponents are lower than the ones in the absence of this interaction [column 2]. This is because the presence of the hydrodynamic interaction couples the dynamics of different residues on the chain. As a result, the non freely-draining (Zimm) chain diffuses faster and has weaker dependence on $N$ compared to the freely-draining (Rouse) chain. 

\subsection{Mean time of contact formation with energy-transfer sink}

The mean time of contact formation, in general, is a sum of reaction and diffusion-controlled parts given by Eqs. (\ref{keq1}) and (\ref{kcorr1}) respectively.  While the integral in Eq. (\ref{keq1}) can be evaluated analytically, there is no closed form analytical expression for Eq. (\ref{kcorr1}).  We, thus, compute Eqs.  (\ref{keq1}) and (\ref{kcorr1}) numerically and substitute the result into Eq. ({\ref{skin}) to obtain  $\tau_{CF}$.  Before carrying out the time-integral in Eq. (\ref{kcorr1}), we non-dimensionlize time in terms of  the time scale of intrinsic quenching $(k_0^{ET})^{-1}$ by defining $t_1 = t k_0^{ET}$. 
\begin{eqnarray}\label{kcorr2}
\tau_D^0 =&=&  \int_0^{\infty} dt_1 \left[\frac{\left< k(t_1) k(0)\right>_{eq}}{\left< k \right>_{eq}^2} - 1 \right] 
\end{eqnarray}
where  $\tau_D^0 = k_0^{ET} \tau_D$ is the dimensionless mean time of diffusion-controlled reaction. In terms of the dimensionless time $t_1$, the modified expression for $\phi(t_1)$ is given by
\begin{equation}
\phi(t_1) = \frac{\sum_{p, odd} \frac{\exp(- p^{\gamma} t_1 /N^{\gamma} k_{eff})}{p^{2\nu + 1}}}{\sum_{p, odd} \frac{1}{p^{2\nu + 1}}},
\end{equation}
where $k_{eff} = k_0^{ET} \tau_0$ is the dimensionless effective rate constant,  $\tau_0$ is the relaxation time $\tau_0 \sim \eta b^3/k_B T$ of the coarse-grained residue of size $b$, $\gamma = 2\nu+1$ and $\gamma = 3\nu$ for the Rouse and Zimm chain respectively. This non-dimensionalizes the time scale of contact formation $\tau_{CF}^0 = k_0^{ET} \tau_{CF}$ such as  $\tau_{CF}^0 = \tau_{R}^0 +  \tau_{D}^0$.  The latter makes the integrals in  Eqs. (\ref{keq1}) and (\ref{kcorr2}) dependent on $k_0^{ET}$ only through the dimensionless effective rate constant $k_{eff}$.

\begin{figure}[t]
\centering
\includegraphics[trim=0.0cm 2.5cm 0cm 1cm,clip=true,scale=0.7]{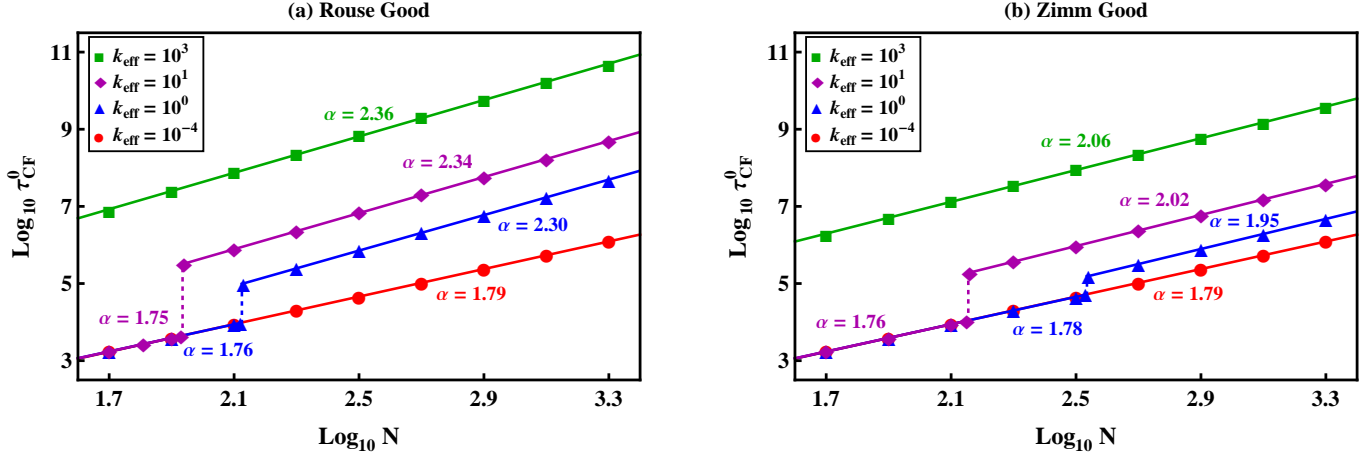}
\caption{Log-log plot for the variation of the dimensionless mean time of contact formation $\tau_{CF}^0$ with $N$ for a chain in a good solvent ($\nu = 3/5$). The diffusion-reaction equation [Eq. (\ref{dre})] is solved in the presence of the exponential sink [Eq. (\ref{etr})] for the different values of the effective (dimensionless) rate constant $k_{eff} = k_0^{ET} \tau_0$ in the (a) absence (Rouse) and (b) presence (Zimm) of hydrodynamic interaction.  Symbols are theoretical results and the solid lines are the corresponding linear fits showing the power law dependence of mean time of contact formation with respect to the number of residues, $\tau_{CF}^0 \sim N^{\alpha}$. The limits of $k_{eff} \ll 1$ and $k_{eff} \gg 1$ show  dominating influence of the reaction-controlled and diffusion-controlled kinetics respectively. In the intermediate limit,  $k_{eff} \approx 1$, the increase in the number of residues switches the kinetics from reaction-controlled at low $N$ to diffusion-controlled at large $N$.  }
\end{figure}


Figs. (2)-(4) show the log-log plot for the variation of $\tau_{CF}^0$ with $N$ in good, theta and poor solvents corresponding to $\nu = 3/5$, $1/2$ and $1/3$ for different values of $k_{eff}$. For typical values of $k_0^{ET} \approx 10^6 -10^9 s^{-1}$ considered in experiments, the dynamics of contact formation has a very sensitive dependence on solvent viscosity. For a fixed value of $k_0^{ET}$, the solvent viscosity is varied by choosing different values of the dimensionless effective rate constant $k_{eff} = k_0^{ET} \tau_0$. Figs. (2)-(4) show that the conditions $k_{eff} \ll 1$ and $k_{eff} \gg 1$ correspond to the dominating influence of the reaction-controlled and diffusion-controlled kinetics respectively. In the former case when the solvent viscosity is low, $k_0^{ET} \ll 1/\tau_0$, the rate of molecular relaxation is faster than rate of reaction between donor and acceptor groups resulting in a reaction controlled kinetics. In the limit of high solvent viscosity, $k_0^{ET} \gg 1/\tau_0$, the contact formation kinetics is determined by the slower rate of molecular relaxation.

In between these limits, $k_0^{ET} \approx 1/\tau_0$, the mean time of contact formation  has significant contributions from both $\tau_R^0$ and $\tau_D^0$. For small $N$, the end-to-end distance correlation relaxes to its equilibrium value fast, making it a reaction controlled process. At large $N$, the process is determined by the diffusion-controlled slow relaxation of the end-to-end distance correlation. The contact dynamics, thus, crosses over from the reaction-controlled kinetics at low $N$ to the diffusion-controlled kinetics at large $N$. This is shown in Figs. (2)-(4) for good, theta and poor solvents in the absence (Rouse) and presence (Zimm) of hydrodynamic interaction. Lower the value of $k_{eff}$, higher the value of $N$ at which the crossover from the reaction-controlled to diffusion-controlled kinetics occurs.

The values of the scaling exponents for the reaction and diffusion-controlled limits are indicated in Figs. (2)-(4) for good, theta and poor solvents respectively. The mean contact formation time for the reaction-controlled kinetics depends on the equilibrium probability distribution. As a result, the scaling exponents for the reaction-controlled kinetics in Figs. (2)-(4) are close to the values predicted by $\tau_{CF}^0 \simeq  N^{3\nu}$. The latter can be obtained by integrating Eq. (\ref{keq1}) analytically followed by an asymptotic limit of large $N$. The exact numerical integration in Eq. (\ref{keq1}), however, yield the values of scaling exponents which are slightly lower then $3\nu$.

\begin{figure}[t]
\centering
\includegraphics[trim=0.0cm 2.5cm 0cm 1cm,clip=true,scale=0.7]{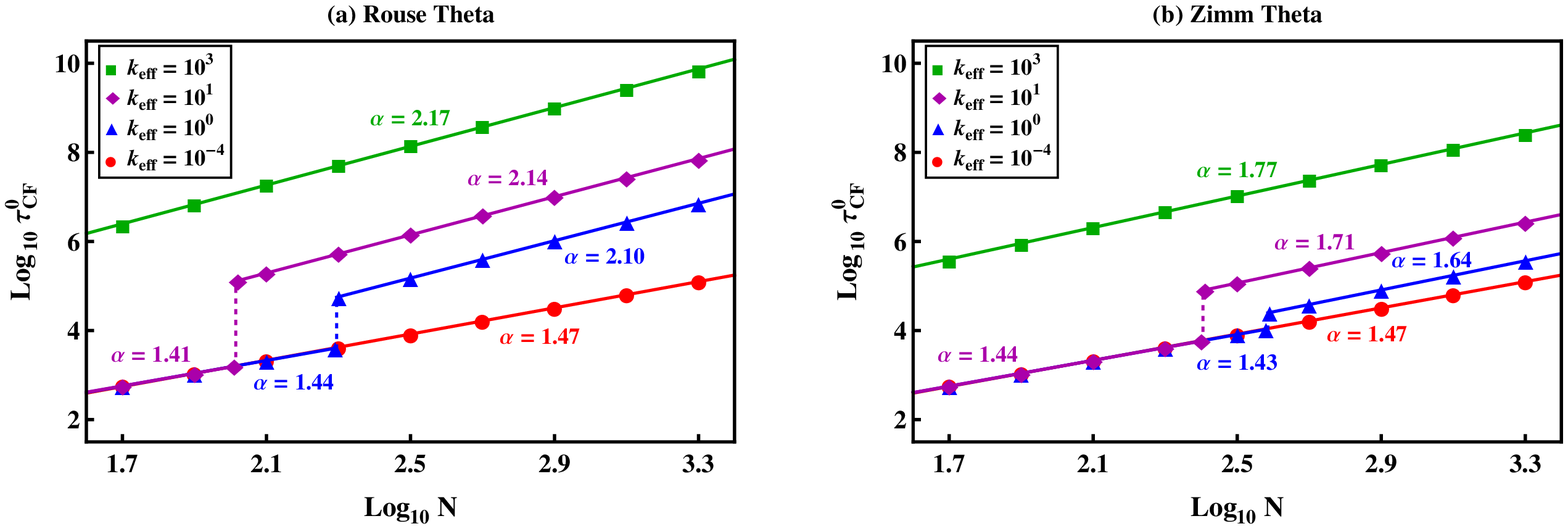}
\caption{Log-log plot for the variation of the dimensionless mean time of contact formation $\tau_{CF}^0$ with $N$ for a chain in a theta solvent ($\nu = 1/2$). The diffusion-reaction equation [Eq. (\ref{dre})] is solved in the presence of the exponential sink [Eq. (\ref{etr})] for the different values of the effective (dimensionless) rate constant $k_{eff} = k_0^{ET} \tau_0$ in the (a) absence (Rouse) and (b) presence (Zimm) of hydrodynamic interaction.  Symbols are theoretical results and the solid lines are the corresponding linear fits showing the power law dependence of mean time of contact formation with respect to the number of residues, $\tau_{CF}^0 \sim N^{\alpha}$. The limits of $k_{eff} \ll 1$ and $k_{eff} \gg 1$ show  dominating influence of the reaction-controlled and diffusion-controlled kinetics respectively. In the intermediate limit,  $k_{eff} \approx 1$, the increase in the number of residues switches the kinetics from reaction-controlled at low $N$ to diffusion-controlled at large $N$.}
\end{figure}

\begin{figure}[t]
\centering
\includegraphics[trim=0.0cm 2.5cm 0cm 1cm,clip=true,scale=0.7]{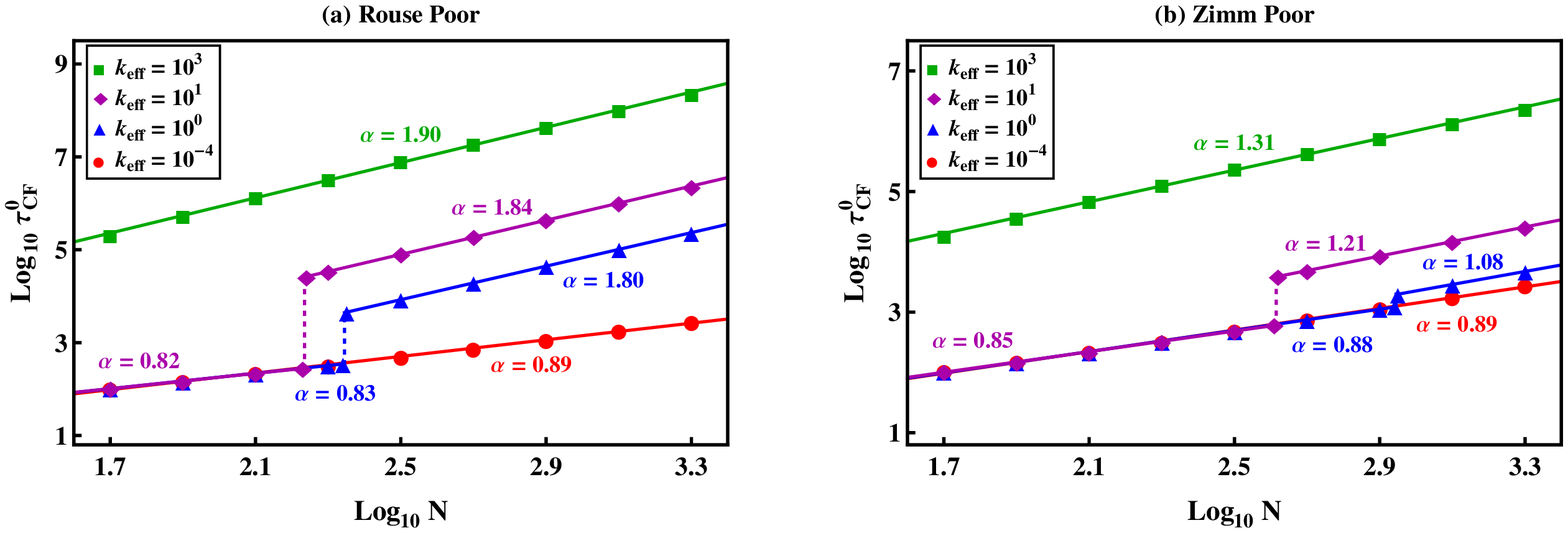}
\caption{Log-log plot for the variation of the dimensionless mean time of contact formation $\tau_{CF}^0$ with $N$ for a chain in a poor solvent ($\nu = 1/3$). The diffusion-reaction equation [Eq. (\ref{dre})] is solved in the presence of the exponential sink [Eq. (\ref{etr})] for the different values of the effective (dimensionless) rate constant $k_{eff} = k_0^{ET} \tau_0$ in the (a) absence (Rouse) and (b) presence (Zimm) of hydrodynamic interaction.  Symbols are theoretical results and the solid lines are the corresponding linear fits showing the power law dependence of mean time of contact formation with respect to the number of residues, $\tau_{CF}^0 \sim N^{\alpha}$. The limits of $k_{eff} \ll 1$ and $k_{eff} \gg 1$ show  dominating influence of the reaction-controlled and diffusion-controlled kinetics respectively. In the intermediate limit,  $k_{eff} \approx 1$, the increase in the number of residues switches the kinetics from reaction-controlled at low $N$ to diffusion-controlled at large $N$.}
\end{figure}

In the limit of $k_{eff} \gg 1$, the dimensionless mean time of contact formation $\tau_{CF}^0$ has dominating influence from the diffusion-controlled kinetics. In this limit, Figs. (2)-(4) show that the values of the scaling exponents in the presence of the exponential sink are slightly higher than the heaviside sink [Table 1]. This is because the scaling exponents for the heaviside sink have been obtained in the sole presence of the diffusion-controlled kinetics [Table 1]. For reaction-diffusion kinetics considered here, the limit of $k_{eff} \gg 1$ ensures small but non-zero contribution from the reaction-controlled part with weaker dependence on $N$. The presence of latter modifies the values of slopes and intercepts in Figs. (2)-(4), resulting in slightly higher values of the scaling exponents.

In the limit of $k_{eff}  \approx 1$, as the solvent quality is varied from good [Fig. 2] or theta [Fig. 3] to poor [Fig. 4], the reaction-controlled kinetics occurs for a larger range of $N$. This is because in a poor solvent the difference in the mean contact formation time for the reaction-controlled and diffusion controlled limits is relatively smaller compared to theta and good solvent conditions. This requires higher values of $N$ to attain the crossover. Similar trend is observed in the presence of the hydrodynamic interaction [Figs. (2b)-(4b)]. The reaction-controlled limit occurs for a higher range of $N$ in the presence of hydrodynamic interaction than in its absence [Figs. (2a)-(4a)]. Again, this is because  in the presence of the hydrodynamic interaction the diffusion-controlled mean time of contact formation are relatively smaller requiring higher values of $N$ to attain the crossover from the reaction-controlled to diffusion-controlled limits.

\subsection{Comparison with experiments}

The end-to-end contact formation dynamics, as measured in different experiments, shows a power law dependence of the mean contact formation time with respect to the number of repeating units on a polypeptide chain with $N \simeq 10-20$. For different polypeptide-solvent systems, the scaling relations have been found to be   $1.05\pm0.06$,\cite{Hudgins2002} $1.5$\cite{Lapidus2002,Buscaglia2003} and $1.7\pm0.1$\cite{Krieger2003, Fierz2005} (good solvent). The scaling exponent of $1.5$ has been rationalized using the SSS description of the diffusion-controlled contact formation dynamics in a theta solvent, which yields $\tau_{sss} \sim N^{3/2}$.\cite{Szabo1980} The SSS prediction, however, deviates from several simulations\cite{Pastor1996, Sakata1976, Podtelezhnikov1997, Chen2005} that predict $\tau_{CF}^d \sim N^2$, in consistent with the WF scaling.\cite{Wilemski1974} The mean time of contact formation in the SSS theory is given by $\tau_{sss} \approx N^{3/2} b^3/D_0 a$, where $D_0 = k_B T/ \zeta$ is the monomer diffusion coefficient and $\zeta$ is the friction coefficient. In a recent study on diffusion-controlled contact formation dynamics, it has been shown that the SSS theory can yield the same result as the WF method if the monomer diffusion coefficient is replaced by the effective diffusion coefficient that accounts for end-to-end relaxation dynamics, thereby accounting for the higher-order modes of the chain.\cite{Toan2008} The latter yields the effective diffusion coefficient as $D_{e} \sim N^{-1/2}$, resulting in the extended SSS prediction $\tau_{esss} \sim N^2$, in agreement with the WF prediction. This implies that the experimentally observed scaling exponent of $1.5$ can be rationalized using the SSS description provided the value of of $D_0$ is considered to be much less than the monomer diffusion coefficient.\cite{Lapidus2002,Buscaglia2003} Thus, the extended SSS prediction of $\tau_{esss} \sim N^2$ is in agreement with other theories and simulations,  but can not rationalize $\tau_{CF} \sim N^{3/2}$ scaling observed experimentally. 

While the above description in terms of the diffusion-controlled  kinetics  of contact formation assumes that the limit of $k_0 \rightarrow \infty$ is always satisfied, the typical values of $k_0^{ET}$ in different experiments lie between $10^6-10^9 s^{-1}$.\cite{Lapidus2000} Based on the present analysis, the dominance of reaction or diffusion-controlled kinetics can be obtained by comparing  $(k_0^{ET})^{-1}$ with the time scale of molecular relaxation $\tau_0 = \eta b^3/k_B T$.  For typical values of $\eta \approx 1-100 cP$,\cite{Lapidus2000} $T \approx 300 K$ and the coarse grained length of  $b \approx 2-5 \AA$, an order of magnitude estimate yields $k_{eff} \approx 10^{-4}-10^{-2}$ for $k_0^{ET} \approx 10^6 s^{-1}$ and $k_{eff} \approx 0.1-10$ for $k_0^{ET} \approx 10^9 s^{-1}$. While the former ($k_{eff} \ll 1$) corresponds to the limit where the reaction-controlled kinetics dominate, the latter ($k_{eff} \approx 1$) points to the crossover region where the reaction-controlled kinetics dominate at low $N$ and the diffusion-controlled kinetics dominate at high $N$. The scaling exponents for the reaction-controlled kinetics obtained here are given by $0.89$, $1.47$ and $1.79$ in poor, theta and good solvents respectively. The fact that they are close to the experimentally determined scaling exponents of ${1.05\pm0.06}$, ${1.5}$ and ${1.7\pm0.1}$ points to the relevance of solvent quality dependence of the mean time of contact formation in the reaction-controlled limit. The importance of the reaction-controlled kinetics in rationalizing the values of the experimentally observed scaling exponents has been proposed earlier.\cite{Yeh2002} 

Compared to the reaction-controlled kinetics, the dominating influence of the diffusion-controlled kinetics ($k_{eff} \gg 1$) yields a relatively stronger dependence of the mean contact formation on $N$. Figs. (2)-(4) show that in good, theta and poor solvents respectively, the freely-draining chains have mean time of contact formation governed by the scaling exponents $2.36$, $2.17$ and $1.90$. For a non-freely-draining chain, in contrast, the respective values of the scaling exponents $2.06$, $1.77$ and $1.31$  in good, theta and poor solvents show  comparatively weaker dependence on $N$.

\section{Conclusions}

Starting from a non-Markovian reaction-diffusion equation that describes the time evolution of distance between two residues on a chain, we have calculated the mean time of end-to-end contact formation using the WF closure approximation. This approach allows us to include the effects of solvent quality and hydrodynamic interaction in a mean-field fashion and yields a power law scaling of the mean contact formation with respect to the number of residues $\tau_{CF} \sim N^{\alpha}$. In the presence of the heaviside sink, the diffusion-controlled mean time of contact formation of a freely-draining chain in good, theta and poor solvents obtained from the present approach are in excellent agreement with the previous theoretical work based on a generalized random walk description. 

The non-Markovian reaction-diffusion equation when supplemented with a more realistic energy transfer sink shows that the interplay of reaction and diffusion-controlled kinetics determine the  mean time of contact formation. In particular, the contact formation dynamics is governed by two time scales, the reciprocal of the intrinsic rate of quenching $(k_0^{ET})^{-1}$, and the relaxation time $\tau_0 = \eta b^3/k_B T$ of the coarse-grained residue of an effective size $b$.  The limits of $k_0^{ET} \tau_0 \ll 1$ and $k_0^{ET} \tau_0 \gg 1$ show  dominating influence of the reaction-controlled and diffusion-controlled kinetics respectively. In the intermediate limit,  $k_0^{ET} \tau_0 \approx 1$, the increase in the number of residues switches the kinetics from reaction-controlled at low $N$ to diffusion-controlled at large $N$.  These results show that the scaling exponent $\alpha$ has sensitive dependence on solvent quality mediated effective interaction between different residues on the protein chain, the solvent viscosity, the hydrodynamic interaction and the magnitude of the intrinsic quenching rate $k_0^{ET}$. From these general results, we conclude that the experimental estimates of the scaling exponent $\alpha$ reflect solvent-quality dependence of the mean contact formation time in the reaction-controlled limit.

\pagebreak

\appendix
\section{Solvent quality dependence of the time correlation of distance fluctuations}

The expression for the time-correlation of distance fluctuations in the absence and presence of hydrodynamics for a chain in a theta solvent has been derived in Ref. (28) using the Rouse and Zimm dynamics respectively. The latter also considers the case of a non-freely draining Zimm chain in a good solvent. Below, we extend this formalism to calculate the dimensionless time correlation of distance fluctuations [Eq. (\ref{corr1})] for a freely draining (Rouse) and non-freely draining (Zimm) chain in good, theta and poor solvents. 

The dynamics of the nth residue at position ${\bf r}_n(t)$ is described by the following Langevin equation\cite{Doi1986}:
\begin{equation}{\label{lang}}
\frac{\partial{\bf r}_n(t)}{\partial t} = \sum _{m} {\bf H}_{nm} \cdot \left( -\frac{\partial U} {\partial {\bf r}_m} + {\bf f}_m(t)\right), 
\end{equation}
where ${\bf H}_{nm}$ is the mobility matrix, $U = U_{el} + U_{int}$ is the interaction potential which accounts for the entropic elasticity due to chain connectivity $U_{el} = \frac{3k_B T}{b^2} \sum_{n=2}^N (r_n - r_{n-1})^2$ and solvent quality dependent effective interaction between chain residues, $U_{int}$. In a theta solvent $U_{int} \approx 0$. In addition, the absence of the hydrodynamic interaction implies that the mobility matrix takes the form ${\bf H}_{nm} = \frac{\bf I}{\zeta} \delta_{nm}$.  The dynamics of a freely draining chain in a theta solvent is, thus, given by the Rouse dynamics 
\begin{equation}
\zeta\frac{\partial{\bf r}_n(t)}{\partial t} =\frac{3k_B T}{b^2}\frac{\partial^2{\bf r}_n(t)}{\partial n^2} + {\bf f}_n(t), 
\end{equation}
where $\zeta$ is the friction coefficient and ${\bf f}_n(t)$ is the random force that follows the Gaussian statistics: $\left<{\bf f}_n(t)\right> = 0$ and $\left<{\bf f}_n(t)\cdot {\bf f}_m(t')\right> = 6 k_B T \delta_{nm} \delta(t-t')$. The condition of free chain ends require ${\partial {\bf r}_n}/{\partial n} = 0$ at $n=0$ and $n=N$. Given this constraint,  the position vector ${\bf r}_n(t)$ in terms of the normal modes can be written as
\begin{equation}\label{dvec}
{\bf r}_n(t) = {\bf X}_0(t) + 2 \sum_{p=1}^{\infty} {\bf X}_p(t) \cos(p\pi n/N) ,
\end{equation}
where $X_0 = \frac{1}{N} \int_0^N dn {\bf r}_n$.  In terms of the normal modes, the Rouse equation is given by
\begin{equation}\label{pnm}
\zeta_p\frac{d{\bf X}_{p}(t)}{d t} = - k_p{\bf X}_p(t)+  {\bf f}_{p}(t), 
\end{equation}
where ${\bf X}_p(t) = \frac{1}{N} \int_0^N ~dn~cos(p \pi n/N) {\bf r}_n(t)$. The above equation can be easily solved to yield
\begin{equation}\label{corrd}
\left<{\bf X}_p(t)\cdot {\bf X}_q(0)\right> = \delta_{pq}\frac{3 k_B T}{k_p} \exp(- t/\tau_p). 
\end{equation}
where $\tau_{p} = \zeta_p/k_p$ is the relaxation time of the pth mode. In a theta solvent $\tau_p = \tau_1/p^2$, where $\tau_1^R = \zeta N^2 b^2/3\pi^2 k_B T$  and $k_p = 6 \pi^2 k_B T p^2/N b^2$. The Rouse dynamics describes the dynamics of a freely-draining chain in a theta solvent.  The effective attractive or repulsive interaction between chain residues in poor or good solvents respectively can be included in the parameter $k_p$ (in a mean field fashion) by determining $k_p = \frac{3 k_B T}{\left<X_p^2\right>_{eq}}$. The latter yields  the solvent quality dependence of $k_p \simeq  p^{1+2 \nu} k_B T/N^{2\nu} b^2$ and $\tau_p^R = \tau_1^R/p^{1+2 \nu}$.\cite{Doi1986}

In the presence of hydrodynamic interaction ${\bf H}_{nn} = \frac{\bf I}{\zeta}$ and ${\bf H}_{nm} = \frac{1}{8\pi\eta_s \left|{\bf R}_{nm}\right|} \left[\hat{{\bf R}_{nm}}\hat{{\bf R}_{nm}} + {\bf I} \right]$ for $n \neq m$. The nonlinear dependence of ${\bf H}_{nm}$ on ${\bf R}_{nm}$ makes the solution of Eq. (\ref{lang}) difficult. The preaveraging approximation of Zimm replaces ${\bf H}_{nm}$ with $\left<{\bf H}_{nm}\right>_{eq} = \int d{\bf R}_{nm} {\bf H}_{nm} P_{eq}({\bf R}_{nm})$, where $P_{eq}$ is given by Eq. (\ref{ssgaussian}). This yields  $\left<{\bf H}_{nm}\right>_{eq} \approx {\bf I}/(|n-m|^\nu \eta_s b) \equiv h(n-m) {\bf I}$. In terms of the normal modes Eq. ({\ref{lang}) can be written as 
\begin{equation}\label{pnmh}
\frac{d{\bf X}_{p}(t)}{d t} =  \sum_q h_{pq} (-k_q{\bf X}_q(t)+  {\bf f}_{q}(t)). 
\end{equation}
Following the linearization approximation\cite{Doi1986}, it can be shown that $\zeta_p = (h_{pp})^{-1} \approx \eta_s b N^{\nu} p^{1-\nu}$. Given that the solvent quality dependence of $k_p$ in the Rouse and Zimm dynamics is the same, the relaxation time of a non-freely draining Zimm chain is given by $\tau_p^Z \simeq \tau_1^Z/p^{3\nu}$, where $\tau_1^Z \simeq N^{3\nu} b^3 \eta_s/k_B T$. This implies that the time correlation of $\left<{\bf X}_p(t)\cdot {\bf X}_q(0)\right>$ in the presence of the hydrodynamic interaction obtained from Eq. (\ref{pnmh}) yield the same expression as Eq. (\ref{corrd}) with $\tau_p$ replaced by $\tau_p^Z$.

Since ${\bf R}_{mn}(t) = {\bf r}_n(t) - {\bf r}_m(t)$, it can be easily shown from Eq. (\ref{dvec}) that
\begin{eqnarray}\label{corrdf1}
\phi_{mn} (t) = \frac{\left< {\bf R}_{mn}(t) \cdot{\bf R}_{mn}(0) \right>}{ \left< {\bf R}_{mn}(0) \cdot {\bf R}_{mn}(0) \right>} &=&  \frac{\sum_{p=1}^{\infty}  \left< {\bf X}_{p}(t) \cdot {\bf X}_{p}(0) \right> [\cos(p\pi n/N) - \cos(p\pi m/N)]^2}{\sum_{p=1}^{\infty}  \left< {\bf X}_{p}(0) \cdot {\bf X}_{p}(0) \right> [\cos(p\pi n/N) - \cos(p\pi m/N)]^2} \nonumber\\
&=& \frac{\sum_{p=1}^{\infty} [\cos(p\pi n/N) - \cos(p\pi m/N)]^2 e^{- t/\tau_p}/p^{1+2 \nu}}{\sum_{p=1}^{\infty} [\cos(p\pi n/N) - \cos(p\pi m/N)]^2/p^{1+2 \nu}},
\end{eqnarray}
which yields Eq. (\ref{corr1}).

\end{document}